\documentclass[footinbib,a4paper,aps,prl,reprint,twocolumn,preprintnumbers,amsmath,amssymb,nobalancelastpage,10pt]{revtex4-1}
\usepackage{amsmath}
\usepackage{verbatim}
\usepackage{graphicx}
\usepackage{color}
\usepackage{tikz}
\usepackage[colorlinks=true,citecolor=blue,linkcolor=blue,urlcolor=blue]{hyperref}

\begin{document}


\title{Strongly Correlated Bosons on a Dynamical Lattice}

\author{Daniel Gonz\'{a}lez-Cuadra,$^1$ Przemys{\l}aw R. Grzybowski,$^{1,2}$ Alexandre Dauphin$^1$ and Maciej Lewenstein$^{1,3}$}

\address{$^1$ICFO-Institut de Ci\`encies Fot\`oniques, The Barcelona Institute of Science and Technology, Av. Carl Friedrich
Gauss 3, 08860 Barcelona, Spain}
\address{$^2$ Faculty of Physics, Adam Mickiewicz University, Umultowska 85, 
61-614 Pozna{\'n}, Poland}
\address{$^3$ICREA, Passeig Lluis Companys 23, 08010 Barcelona, Spain}

\begin{abstract}

We study a one-dimensional system of strongly-correlated bosons on a dynamical lattice. To this end, we extend the standard Bose-Hubbard Hamiltonian to include extra degrees of freedom on the bonds of the lattice. We show that this minimal model exhibits phenomena reminiscent of fermion-phonon models. In particular, we discover a bosonic analog of the Peierls transition, where the translational symmetry of the underlying lattice is spontaneously broken. This provides a dynamical mechanism to obtain a topological insulator in the presence of interactions, analogous to the Su-Schrieffer-Heeger (SSH) model for electrons. We characterize the phase diagram numerically, showing different types of bond order waves and topological solitons. Finally, we study the possibility of implementing the model using atomic systems.

\end{abstract}

\maketitle

\paragraph{Introduction -- }

The study of interactions between particles and lattice degrees of freedom is of central importance in quantum many-body physics.  The interplay between electrons and phonons has been extensively studied, leading to the description of paradigmatic effects such as superconductivity, polaron formation or charge density waves \cite{altland,emin}. The analogous problem for bosons, on the other hand, has not been extensively investigated. 
The basic feature of phononic systems is that the lattice may fluctuate or order at various wavelenghts.
In one dimension, a system of itinerant particles on a deformable lattice can undergo a Peierls transition \cite{peierls}, characterized by the spontaneous breaking of the lattice translational symmetry in a density-dependent manner.
For fermions, the statistical correlations induced by Pauli's exclusion principle are sufficient to drive this effect, associated with a gap opening around the Fermi surface. The latter is absent in the bosonic case. However, similar effects still appear in the presence of sufficiently strong interactions, as we report in this work.

The study of boson-lattice problems becomes very relevant in the context of quantum simulators. These are versatile platforms where model Hamiltonians can be engineered with an unprecedented degree of control~\cite{quantum_simulation_2,quantum_simulation_1}. Ultracold atoms in optical lattices, in particular, allow one to experimentally address systems of strongly-correlated bosons and to study their properties~\cite{jaksch,bloch,bloch_2,maciej}, e.g.  the realization of the phase transition between a Mott insulator and a superfluid~\cite{superfluid_mott} in the Bose-Hubbard model~\cite{fisher}. Since then, a variety of models have been studied\textemdash including, e.g. different types of interactions~\cite{eBH_1,non_standard_bose_hubbard} or artificial gauge fields~\cite{artificial_gauge_1,artificial_gauge_2}. These provide an interesting platform to study novel phenomena, such as supersolid phases~\cite{eBH_2} or topological order~\cite{haldane_insulator}.

The simulation of these models rely on the implementation of static optical lattices. The particles do not influence the lattice structure and, therefore, phonons are usually not taken into account. Trapped ion systems can also simulate many-body Hamiltonians \cite{blatt,quantum_simulation_ions_1,blatt,quantum_simulation_ions_2}. In these systems, phonons appear naturally \cite{wineland}, and can be used to mediate interactions between the ions \cite{cirac_porras}. However, trapped ions are confined at the lattice sites, making the simulation of itinerant particles more challenging. Recently, advances in designing systems formed by both neutral atoms and ions \cite{state_dependent_hopping,ions_atoms_1,ions_atoms_2,ions_atoms_3, ions_atoms_4,ion_atoms_exp_1,ion_atoms_exp_2,ion_atoms_exp_3} suggest the possibility of simulating itinerant particles and dynamical lattices simultaneously. This strategy was explored for a chain of fermionic atoms, where a Peierls transition was predicted \cite{ions_atoms_1}. Alternative approaches include the use of molecules in self-assembled dipolar lattices \cite{dipolar_molecules}, optical cavities \cite{cavity_1,cavity_2,cavity_3,cavity_4,cavity_5,cavity_6} or trapped nanoparticles \cite{nano}.

In this letter, we propose and analyze a one-dimensional model of interacting bosons coupled to a dynamical lattice, i.e. deformable and non-adiabatic. We also discuss a possible experimental scheme with ultracold atoms. The most important result is the discovery of bosonic analogs of the Peierls transition, leading to commensurate and incommensurate Bond Order Waves (BOW). For density $\rho=1/2$, in particular, the ground state corresponds to a dynamically-generated topological insulator, supporting edge states and topological solitons, similar to the fermionic SSH model \cite{ssh_polymers}. The proposed model provides a unique playground to study the interplay between strong interactions, lattice dynamics, spontaneous symmetry breaking and topological effects.

\paragraph{Model -- }

We introduce a minimal model of strongly-correlated bosons interacting with lattice degrees of freedom described by a set of independent two-level systems. The Hamiltonian reads
\begin{equation}
\label{eq:non_standard_ham}
\begin{aligned}
\hat{H}&=-t\sum_i \left(\hat{b}^\dagger_i\hat{b}^{\vphantom{\dagger}}_{i+1}+\text{h.c.}\right)+\frac{U}{2}\sum_i \hat{n}_i(\hat{n}_i-1)-\mu\sum_i\hat{n}_i\\
&-\alpha\sum_i\left(\hat{b}^\dagger_i\hat{\sigma}^z_{i}\hat{b}^{\vphantom{\dagger}}_{i+1}+\text{h.c.}\right)+\frac{\Delta}{2}\sum_i\hat{\sigma}^z_{i}+\beta\sum_i\hat{\sigma}^x_{i}\,,
\end{aligned}
\end{equation}
where $\hat{b}^\dagger_i$ creates a boson on site $i$ and $\hat{n}_i=\hat{b}^\dagger_i\hat{b}^{\vphantom{\dagger}}_i$ is the number operator.  $\hat{\sigma}^z_i$ and $\hat{\sigma}^x_i$ are Pauli operators associated with a spin-1/2 system living on the bond between sites $i$ and $i+1$. The first three terms of (\ref{eq:non_standard_ham}) correspond to the standard Bose-Hubbard Hamiltonian \cite{fisher}. The next term describes a lattice-dependent boson tunneling. The total hopping through a bond is maximized (resp. minimized) for a spin in the ``up" (``down") state. Finally, the last two terms introduce the spin dynamics. The Hamiltonian (\ref{eq:non_standard_ham}) bears similarities with models with spin dependent hoppings, such as quantum link models~\cite{wiese}.

In this work, we focus on the regime of quasi-adiabatic spins ($\beta\ll t$). In this limit, the ground state of the spins depends on the competition between two terms: the energy difference $\Delta$ and the interaction $\alpha$ with the bosons. If one dominates, the expectation value $\langle\hat{\sigma}^{z}_i\rangle$ will be uniform and close to $-1$ ($\Delta\gg \alpha$) or $+1$ ($\Delta\ll\alpha$). When the two are comparable, phases with broken translational symmetry arise. 

\paragraph{Hardcore Bosons --}

Consider an adiabatic lattice, $\beta=0$, in the hardcore boson limit, $U\rightarrow\infty$. After a Jordan-Wigner transformation, the model is mapped to a system of spinless fermions in a classical background. At half filling, the spin configuration minimizing energy is staggered (Neel order) for values of $\Delta$ between two critical points, $\Delta_{c}^{\pm}=\frac{4t}{\pi}[\delta\pm (E(1-\delta^2)-1)]$, where $\delta=\alpha/t$ and $E(x)$ is a complete elliptic integral of the second kind~\footnote{See Supplementary Material for further details on the ground state properties of the model in the hardcore limit, the properties of the BOW phases, the numerical method used and the experimental realization, which includes Refs. \cite{calabrese,area_law,fidelity_susceptibility}}. From the fermions' viewpoint, this leads to the development of a staggered order on the bonds, a gap opens at the Fermi surface, and the system becomes insulating. This effect appears when the lattice deformation, which breaks translational invariance, has a wavelength equal to $\pi/k_F$, where $k_F$ is the Fermi wavevector. This is the mechanism behind the Peierls instability~\cite{peierls}. Thus, our minimal model is capable of describing analogous phenomena such as those appearing in more complicated fermion-lattice systems~\cite{ssh_polymers}.

\begin{figure}[t]
  \centering
  \includegraphics[width=1.\linewidth]{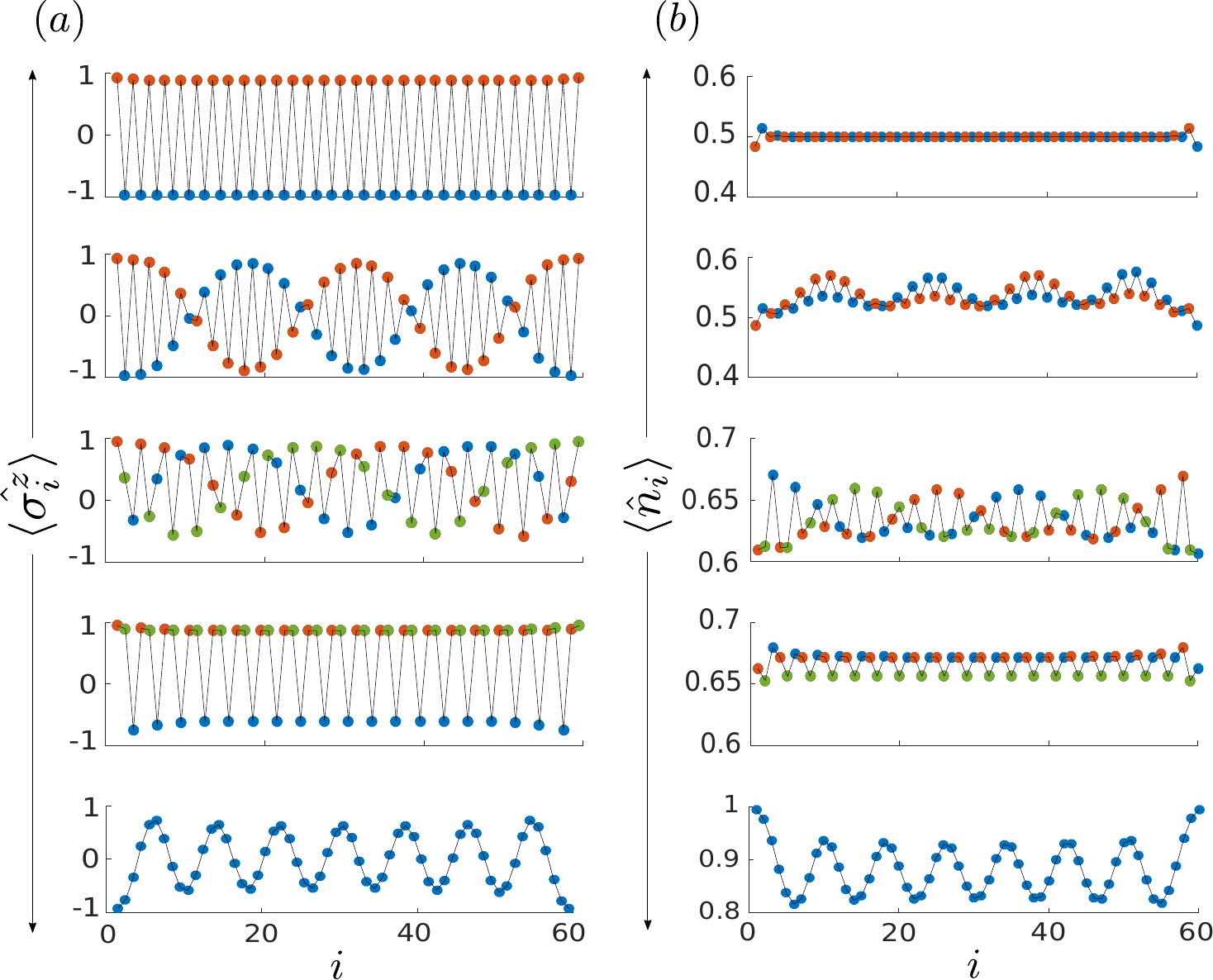}
\caption{\label{fig:figure1} Spatial structure of bond (a) and site (b) expectation values for $\Delta=0.85$ and different bosonic densities, showing some of the representative orders that can develop.
From above to below: $\rho=1/2$, $\rho=1/2+2/60$, $\rho=2/3-2/60$, $\rho=2/3$ and $\rho=0.88$. Different colors represent different sublattice elements, making explicit the long-wavelength modulations on top of the underlying order.}
\end{figure}

\paragraph{Finite Interactions -- }

For finite values of $U$, we enter into the strongly-correlated boson regime, and the mapping to non-interacting fermions is not possible. To calculate the ground state of the system, we use a DMRG algorithm with bond dimension $D=40$ \cite{dmrg}. We consider a system size of $L=60$ sites (and $L-1$ bonds), and work with open boundary conditions. We truncate the maximum number of bosons per site to $n_0=2$. This approximation is justified for low densities and strong interactions~\cite{Note1}. In the following, we fix the values of the parameters to $t=1$, $\alpha=0.5$ and $\beta=0.02$.

At the bosonic density $\rho=1/2$, the Neel order survives for finite values of $U$, and disappears for small interactions. Strong correlations are needed, therefore, to have a bosonic Peierls phase. The Bose-Hubbard model on a fixed bond-dimerized lattice was previously studied, revealing an insulating phase at $\rho=1/2$ \cite{superlattice}, and the presence of topological edge states \cite{ssh_bose_hubbard}. Here, the same superlattice structure is obtained dynamically, in the spirit of the original SSH model for fermions and phonons \cite{ssh_polymers}.  We also observe edge states which will be studied in a separate work~\cite{unpublished}. We focus here on a different topological effect also present in the SSH model: the solitonic solutions. These are a consequence of the double degenerate ground state at $\rho=1/2$, corresponding to the two inverted staggered patterns, and only occur when quantum fluctuations on the lattice are present.

For $U=10$, we study the phase diagram of the model in terms of $\Delta$ and $\rho$. For $\Delta\gg\alpha$ or $\Delta\ll\alpha$, the spin configuration in the ground state is uniform. 
The bosonic part of the Hamiltonian (\ref{eq:non_standard_ham}) is qualitatively similar to the Bose-Hubbard model \cite{fisher}, with a Mott insulator (MI) and a superfluid phase (SF). In an intermediate regime ($\Delta\approx0.6-1.0$), the translational symmetry is broken in the ground state for a substantial range of densities. Figure~\ref{fig:figure1} shows the spatial structure on the bonds (a) and sites (b) for $\Delta=0.85$. 
For $\rho=1/2$ and $2/3$, the unit cell is enlarged to two and three sites, respectively. Similarly to $\rho=2/3$, a trimer configuration appears for $\rho=1/3$ at a different $\Delta$. For densities close to the mentioned ones, long wavelength modulations appear on top of the corresponding patterns. These are solitonic configurations where the underlying order\textemdash staggered in the half-filled case\textemdash is reversed periodically forming kinks;  the ``extra" bosons or holes lead to increased density modulations, located around the kinks (2nd or 3rd row in panel (b)).
Finally, close to $\rho=1$, long wavelength structures appear. The bosonic hopping $\langle\hat{b}^\dagger_i\hat{b}^{\vphantom{\dagger}}_{i+1}+\text{h.c.}\rangle$ presents the same spatial pattern as $\langle\hat{\sigma}^z_i\rangle$ in all the cases. We therefore focus on the latter quantity for simplicity. 
The ground states, shown in Fig.~\ref{fig:figure1}, possess long-range order. We refer to the corresponding quantum phases as \textit{Bond Order Waves} (BOW), since the bosonic order is block-diagonal. In many cases, this bond order is accompanied by small density waves. We consider the spin structure factor

\begin{figure}[t]
  \centering
  \includegraphics[width=0.9\linewidth]{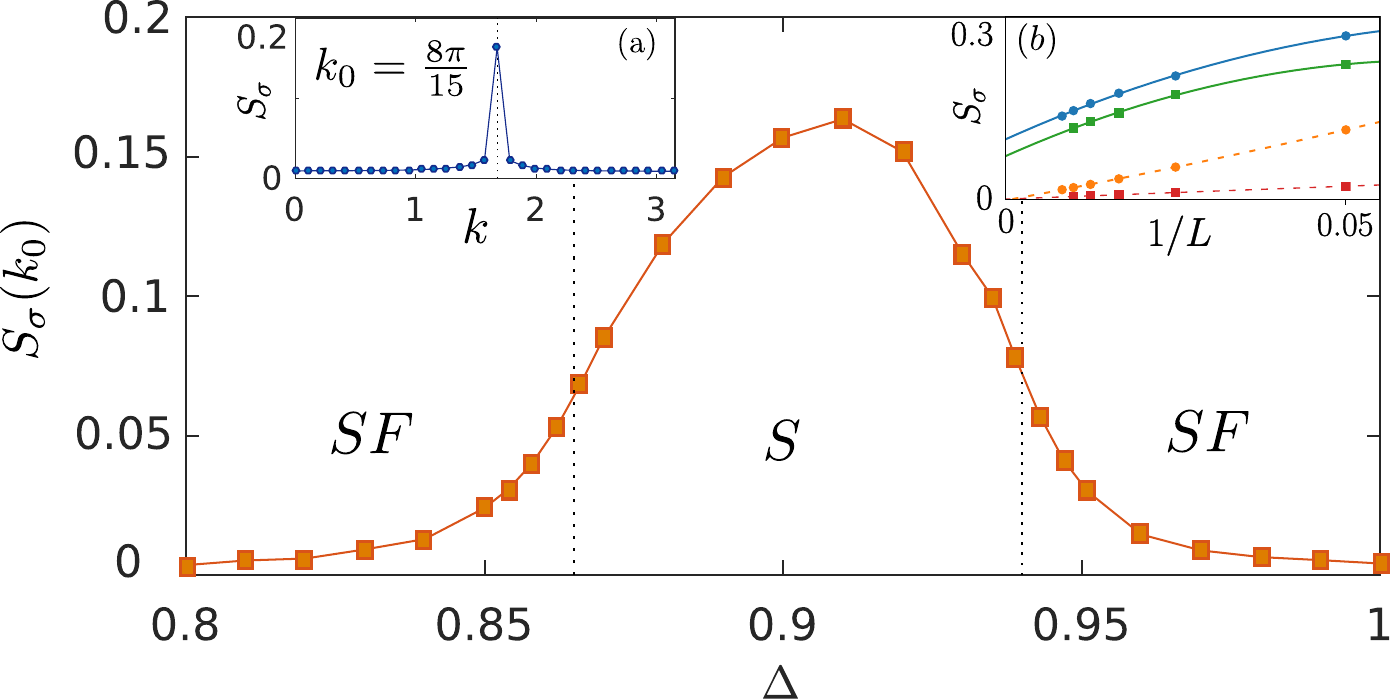}
\caption{Structure factor at $k_0$ in terms of $\Delta$. It allows one to qualitatively distinguish between the iBOW and the  uniform SF phases. The exact location of the critical points (dotted lines) is found through a finite-size scaling of the fidelity susceptibility~\cite{Note1}.
\label{fig:figure2} Inset: (a) Structure factor $S(k)$ in the solitonic phase (S), for $\rho=0.733$ and $\Delta=0.90$. A clear peak is observed at $k_0=8\pi/15$. (b) Finite-size scaling of $S_\sigma(k)$ for $\rho=0.55$ (circles) and $\rho=0.85$ (squares), for $k=k_0(\rho)$ (continuous line) and $k=\pi$ (dashed line).}
\end{figure}

\begin{equation}
\label{eq:structure_factor}
S_\sigma(k)=\frac{1}{L^2}\sum_{i,j} e^{(x_i-x_j)ki}\langle\left(\hat{\sigma}^z_i-\bar{\sigma^z}\right)\left(\hat{\sigma}^z_j-\bar{\sigma^z}\right)\rangle\,,
\end{equation}
with $\bar{\sigma}^z=\sum_i \langle\hat{\sigma}_i\rangle/L$, where the summations run over all bonds. This quantity develops a peak for some $k_0$ in the presence of long-range order, and its height can be used as an order parameter. Figure~\ref{fig:figure2} shows $S_\sigma^{\text{max}}$ in terms of $\Delta$ for $\rho=0.733$, which  qualitatively distinguishes a uniform SF phases from a solitonic BOW. The inset (a) presents $S_\sigma$ for $\Delta=0.90$, where a peak develops for $k_0=8\pi/15$. From this wavevector, an order wavelength can be defined as $\lambda_0=2\pi/k_0$. We note that, in this case, $\lambda_0$ is not an integer factor of the lattice spacing $a$ (fixed to one here). This is also the case for the other solitonic and long-wavelength BOW phases. We refer to these orders as \textit{incommensurate} (iBOW). For $\rho=1/3,\,1/2$ and $2/3$, however, $S_\sigma$ presents a peak at $\pi/3$, $\pi$ and $2\pi/3$, respectively, with wavelengths of the form $a\mathbb{N}$. We call the latter \textit{commensurate} orders (cBOW). While a long-range order is expected in commensurate phases, its presence in incommensurate ones (especially solitonic) is a special feature of the model, related to the Peierls instability. The inset (b) shows the scaling of $S_\sigma(k)$ with the system size, for $k=k_0$ and $k=\pi$ and for two representative incommensurate cases: $\rho=0.55$ (solitonic) and  $\rho=0.85$ (long-wavelength). The fit, containing terms up to $O(1/L^3)$, shows that the long-range order exists in the thermodynamic limit.

\begin{figure}[t]

  \centering
  \includegraphics[width=0.85\linewidth]{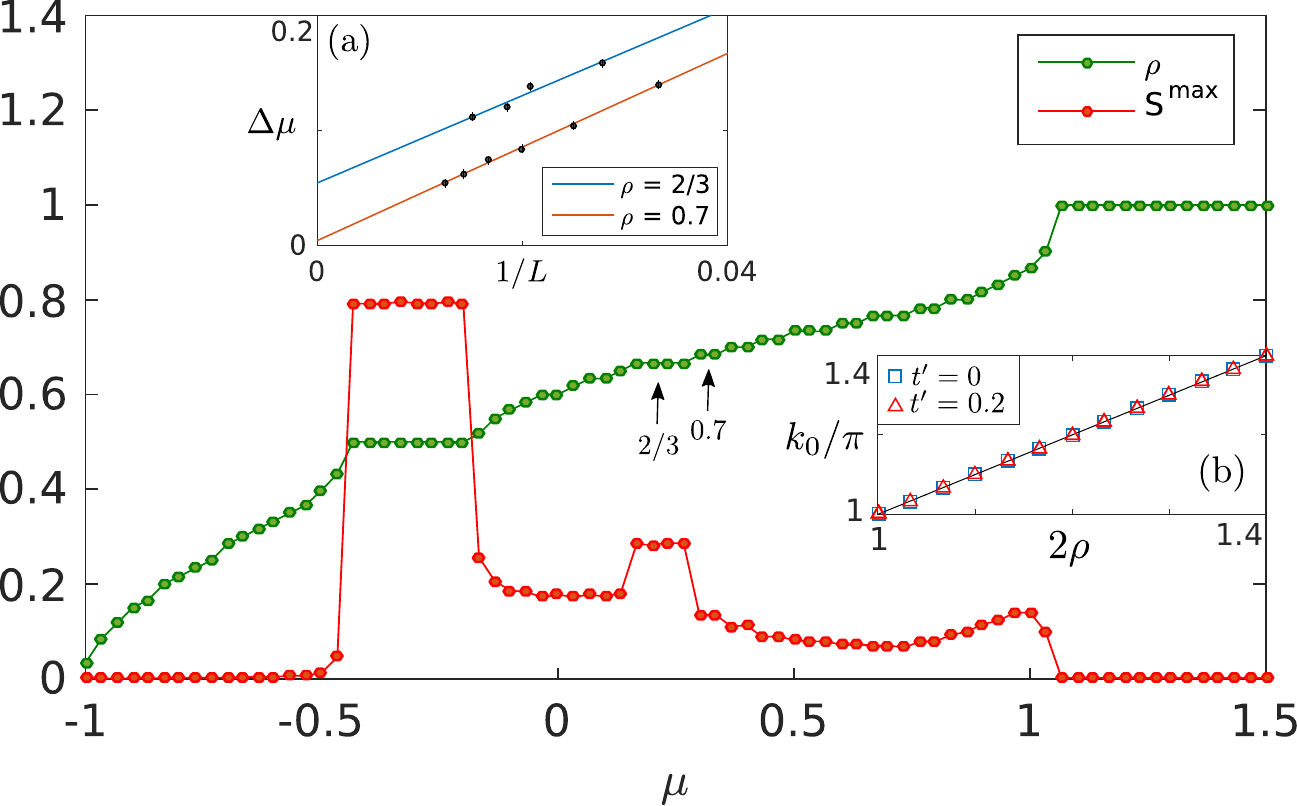}
\caption{\label{fig:figure4} Density $\rho$ and maximum structure factor $S^{max}$ in terms of the chemical potential $\mu$ for $\Delta=0.87$ and $L=60$. The structure factor has non-zero values for the BOW 
phases. Plateaus in the density are related to incompressible phases, but they can also appear as finite size effects. Insets: (a) Scaling of the plateaus $\Delta\mu$ for different system sizes for a cBOW phase ($\rho=2/3$) and for a solitonic iBOW one ($\rho=0.7$) (b) $k_0$ vs $\rho$ for $t^\prime=0$ and $t^\prime=0.2$ (see main text).}
\end{figure}

One of the principal features of the theory of Peierls transition is the relation between the order wavevector and the Fermi wavevector \cite{peierls}. 
In one-dimensional systems with a two-point Fermi surface, the theory predicts $k_0=2 k_F=2\rho \pi$, independently of the fermion dispersion and the form of the fermion-lattice interaction. Remarkably, we found the same relation 
for bosonic Peierls transitions (inset (b) of Fig.~\ref{fig:figure4}), where the Fermi surface is absent. This relation holds in the presence of next neighbor hopping $- t'\sum_i \left(\hat{b}^\dagger_i\hat{b}^{\vphantom{\dagger}}_{i+2}+\text{h.c.}\right)$, where even hard-core bosons cannot be mapped onto fermions~\footnote{We note that the presence of next-nearest neighbor hopping terms changes the topology of the systems from a chain into a zig-zag ladder, which raises the question about similar results in two-dimensional systems.}. This suggests that Peierls transitions require a deeper theory unifying the fermionic and bosonic cases.

There are no off-diagonal bosonic long-range orders coexisting with the BOW order. We found superfluid, on-site pair superfluid and inter-site pair superfluid correlations to decay exponentially in the BOW phases~\cite{Note1}. Additionally, the scaling of the entanglement entropy shows that all BOW phases are gapped, although in the case of iBOW phases the gap is probably quite small~\cite{Note1}. Therefore, the solitonic phases present in our model are qualitatively different from those appearing in the extended Bose-Hubbard model~\cite{eBH_7}.

Interestingly, the iBOW phases are compressible, with compressibility $\kappa=\frac{\partial\rho}{\partial\mu}\neq 0$. This is in contrast to the behavior of many bosonic models, where the presence of a gap and a diagonal or block-diagonal order usually implies incompressibility. Figure~\ref{fig:figure4} depicts the density $\rho$ in terms of $\mu$ for $\Delta=0.87$. Here, a superfluid phase occurs for $0<\rho<1/2$, and BOW phases appear for $1/2\leqslant\rho<1$. Finally, $\rho=1$ corresponds to a MI. The plateaus in the $\mu-\rho$ line signal the incompressible phases, which, apart from the MI, correspond to a cBOW phase at $\rho=1/2$ and $\rho=2/3$. The finite size scaling of other plateaus (inset (a)) reveals that the iBOW phases are indeed compressible. 

In the hardcore limit, the presence of a gap, together with a long-range order, and a non-zero compressibility can be understood using the single-particle fermionic picture. 
For increasing particle density, the added particles will not occupy states above the gap. The  $k_0=2k_F=2\rho \pi$  relation means that the position of the gap will be adjusted to the new Fermi level. The composite spin-particle system avoids the gap penalty by the modification of the effective lattice structure.
For strongly-correlated bosons the mechanism is more complicated, as the Fermi energy picture is lacking. However, many of the properties remain. In particular, the relation $k_0=2k_F=2\rho \pi$ still holds. Therefore, neither the presence of a gap nor long-range order necessarily exclude compressibility. 
Nonetheless, the commensurate orders are incompressible (cBOW).
This implies that these orders are more stable under small changes of the chemical potential. Figure~\ref{fig:figure4} shows the maximum value of the structure factor ($S^{max}_\sigma$) as a function of $\mu$. It is zero for the uniform phases (MI and SF) and it changes continuously among the BOW phases, except for the commensurate orders where it clearly stands out. Since $S^{max}_\sigma$ represents an order parameter, this behavior corresponds to finite changes in the free energy as the density is varied, meaning that these \textit{pinned} wavelengths are energetically more stable. 

\begin{figure}[t]
  \centering
  \includegraphics[width=1.0\linewidth]{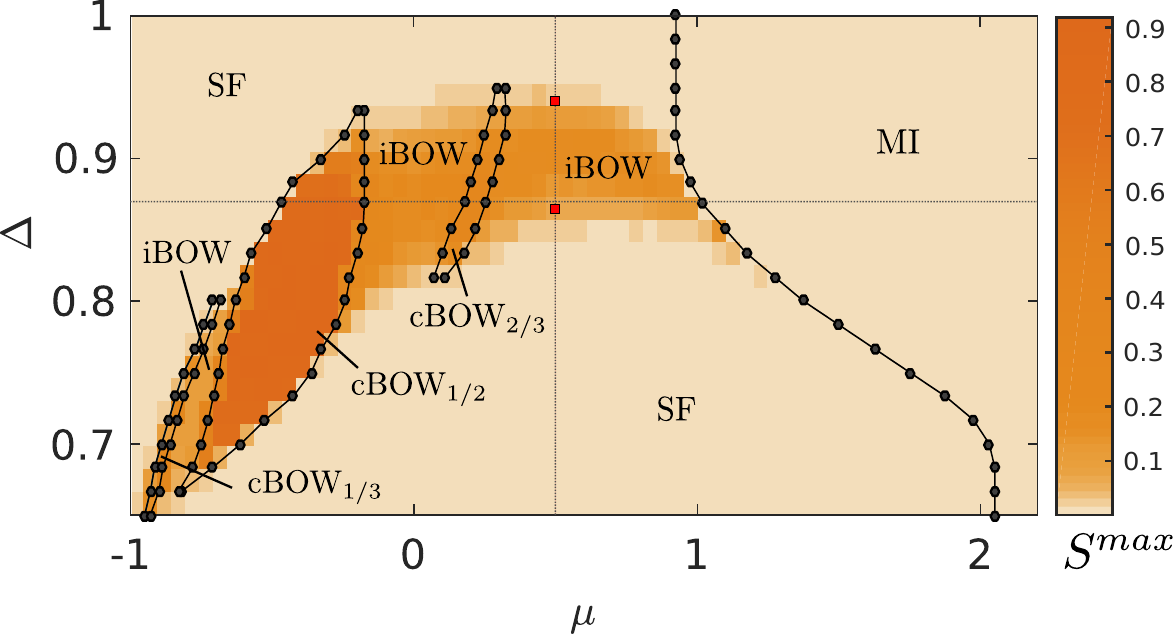}
\caption{\label{fig:figure5}Phase diagram of the Hamiltonian (\ref{eq:non_standard_ham}) for a system size of $L=60$ in terms of $\Delta$ and $\mu$. The solid black lines delimit the incompressible phases (cBOW and MI). The maximum value of the structure factor is represented by the color plot, qualitatively distinguishing between the iBOW and SF phases. The dotted lines correspond to the cuts for $\mu=0.5$ (Fig.~\ref{fig:figure2}) and $\Delta=0.87$ (Fig.~\ref{fig:figure4}). For the former, the red squares mark two critical points in the thermodynamic limit.}
\end{figure}

For a wide range of values of $\Delta$, we calculate the plateau size and the maximum structure factor in terms of $\mu$. These two properties are sufficient to identify all the phases of the model. The results are summarized in the phase diagram  (Fig.~\ref{fig:figure5}). Inside the MI, the spins are uniform and $\langle \hat{\sigma}^z_i\rangle$ changes continuously from $+1$ to $-1$ as $\Delta$ increases. As a consequence, the boundary between this phase and the SF is modified. The phase diagram also shows the extensions of the BOW phases, the most stable one being cBOW$_{1/2}$.

\paragraph{Experimental implementation -- } 

To realize the proposed model \eqref{eq:non_standard_ham}, we consider first a gas of ultracold bosonic atoms in an optical lattice, described by the Bose-Hubbard Hamiltonian. A second optical lattice, trapping either neutral or charged atoms, is introduced, placing its minima between two minima of the first lattice. The atoms corresponding to the second lattice have two internal degrees of freedom\textemdash representing spin systems\textemdash and the potential is deep enough to confine them  \cite{Note1}. As shown in \cite{atom_ion_zoller,ions_atoms_2,atom_ion_many_body,state_dependent_hopping}, in this situation, the hopping of the moving particles between two neighboring sites is influenced by the internal state of the corresponding spin, giving rise to the desired boson-spin interaction. The on-site boson interaction term can be influenced by the internal state of the spins. However, this dependence is very weak \cite{ions_atoms_2}, and we neglect it here. The spin part of the Hamiltonian can be implemented as follows: the energy difference between the two spin states is obtained by introducing an external magnetic field, and the spin flipping is enforced using laser-assisted transitions between the two states. This strategy is valid both when the impurity corresponds to a neutral atom or to an ion. Although the boson-spin interaction $\alpha$ might be difficult to tune in an experiment, the phases we show in this work are present for a broad range of values of this parameter, for a  suitably chosen $\Delta$. The different phases could be detected by measuring the spin structure factor \cite{spin_correlations_1,spin_correlations_2,spin_correlations_3} and the compressibility in the atomic system \cite{compressibility_1,compressibility_2}.

\paragraph{Summary --}

We introduced a boson-spin Hamiltonian that models the behavior of strongly-correlated bosons on a dynamical lattice, and demonstrated the possibility of obtaining bosonic analogs of the Peierls phase. We characterized the phases of the system in the quasi-adiabatic limit (slow lattice dynamics), using the spin structure factor, entanglement entropy and compressibility. We found, besides the uniform SF and MI phases,  compressible and incompressible Bond Order Waves. We also discussed the possibility of implementing the model using ultracold atoms and ions trapped in optical lattices. In the future, it would be interesting to study more extensively the topological properties of the model, as well as the regime of non-adiabatic spins.

\begin{acknowledgments}

The authors thank A. Celi, R. W. Chhajlany, A. Piga, L. Tarruell, and E. Tirrito for useful discussions. This project has received funding from the European Union's Horizon 2020 research and innovation programme under the Marie Sk\l{}odowska-Curie grant agreement No 665884, the Spanish Ministry MINECO (National Plan 15 Grant: FISICATEAMO No. FIS2016-79508-P, SEVERO OCHOA No. SEV-2015-0522, FPI), European Social Fund, Fundació Cellex, Generalitat de Catalunya (AGAUR Grant No. 2017 SGR 1341 and CERCA/Program), ERC AdG OSYRIS, EU FETPRO QUIC, and the National Science Centre, PolandSymfonia Grant No. 2016/20/W/ST4/00314. A. D. is financed by a Cellex-322 ICFO-MPQ fellowship.

\end{acknowledgments}

\bibliographystyle{apsrev4-1}
\bibliography{bibliography_v2}

\end{document}



\title{Supplementary Material to ``Strongly Correlated Bosons on a Dynamical Lattice"}

\author{Daniel Gonz\'{a}lez-Cuadra,$^1$ Przemys{\l}aw R. Grzybowski,$^{1,2}$ Alexandre Dauphin$^1$ and Maciej Lewenstein$^{1,3}$}

\address{$^1$ICFO-Institut de Ci\`encies Fot\`oniques, The Barcelona Institute of Science and Technology, Av. Carl Friedrich
Gauss 3, 08860 Barcelona, Spain}
\address{$^2$ Faculty of Physics, Adam Mickiewicz University, Umultowska 85, 
61-614 Pozna{\'n}, Poland}
\address{$^3$ICREA, Passeig Lluis Companys 23, 08010 Barcelona, Spain}

\begin{abstract}

In this Supplementary Material we discuss additional details concerning the hardcore limit of the boson-spin Hamiltonian (1) introduced in the main text, the properties of the Bond Order Wave phases presented there, the experimental realization of the model and the numerical method used.

\end{abstract}

\maketitle

\section{Hardcore Bosons in a Static Lattice}

Consider the boson-spin Hamiltonian (1) of the main text for a static lattice, with $\beta=0$. In the hardcore boson limit, $U\rightarrow\infty$, a Jordan-Wigner transformation maps the system to a model of spinless fermions. The transformed Hamiltonian is quadratic in the fermionic operators,
\begin{equation}
\begin{aligned}
\label{eq:static_fermion_ham}
\hat{H}=&-t\sum_i \left(\hat{c}^\dagger_i\hat{c}^{\vphantom{\dagger}}_{i+1}+\text{h.c.}\right)-\mu\sum_i\hat{n}_i\\
&-\alpha\sum_i\left(\hat{c}^\dagger_i\hat{\sigma}^z_{i}\hat{c}^{\vphantom{\dagger}}_{i+1}+\text{h.c.}\right)+\frac{\Delta}{2}\sum_i\hat{\sigma}^z_{i}\,,
\end{aligned}
\end{equation}
where $\hat{c}^\dagger_i$ and $\hat{c}_i$ are creation and annihilation fermionic operators, respectively, and $\hat{n}_i=\hat{c}^\dagger_i\hat{c}_i$ is the number of fermions at site $i$. This Hamiltonian (\ref{eq:static_fermion_ham}) describe a system of non-interacting fermions coupled to classical degrees of freedom. For a given configuration of the classical variables, the Hamiltonian can be diagonalized analytically using a single-particle picture.




For large enough values of $\Delta$ the fermionic and spin subsystems decouple. The spin configuration that minimizes the total energy is the one with all spins down. Conversely, for $\Delta=0$ the spin configuration in the ground state of the system is the one that minimizes the energy of the fermion subsystem. For $\rho\notin{\{0,1\}}$, this happens when all spins are in the up state, making the fermion hopping uniform and maximal. For other values of $\Delta$, the energies of these two configurations become comparable and other spin configurations are possible in the ground state.

We focus on the half-filling case  ($\rho=1/2$). In the uniform ``down" and ``up" spin configurations, the ground state energy per site is  $\varepsilon=-\Delta/2-2(t-\alpha)/\pi$ and $\varepsilon=\Delta/2-2(t+\alpha)/\pi$, respectively. There is another important spin configuration, the Neel ordered or staggered spin structure. In this configuration, the values of the fermion hopping are also staggered ($t\pm\alpha$), the unit cell doubles and a gap opens around the Fermi energy. There are two branches of single-particle energies in the reduced Brillouin zone ($-\pi/2<k<\pi/2$),
\begin{equation}
\epsilon_{\pm}(k)=\pm2t\sqrt{\delta^2\sin^2k+\cos^2k}\,,
\end{equation}
where $\delta=\alpha/t$. 
This leads to a ground state energy per site of $\varepsilon=-2 t E(1-\delta^2)/\pi$ where $E(x)$ is complete elliptic integral of the second kind. By comparing the aforementioned energies, we conclude that the staggered spin pattern energy is lower between two critical values of the parameter $\Delta$,
\begin{equation}
\Delta_{c}^{\pm}=\frac{4t}{\pi}[\delta\pm (E(1-\delta^2)-1)].
\end{equation}
On the other hand, the uniform ``down" and uniform ``up" configurations have a lower energy for $\Delta>\Delta_c^{+}$ and $\Delta<\Delta_c^{-}$,respectively. We have checked numerically that, indeed, these configurations correspond to the ground state of the system in the respective regimes, being the only possible ones at half filling.

\section{Bond Order Wave Phases}

In this section, we discuss in more detail some of the properties of the BOW phases presented in the main text.

\subsection{Presence of a Gap}

\begin{figure}[t]
  \centering
  \includegraphics[width=0.9\linewidth]{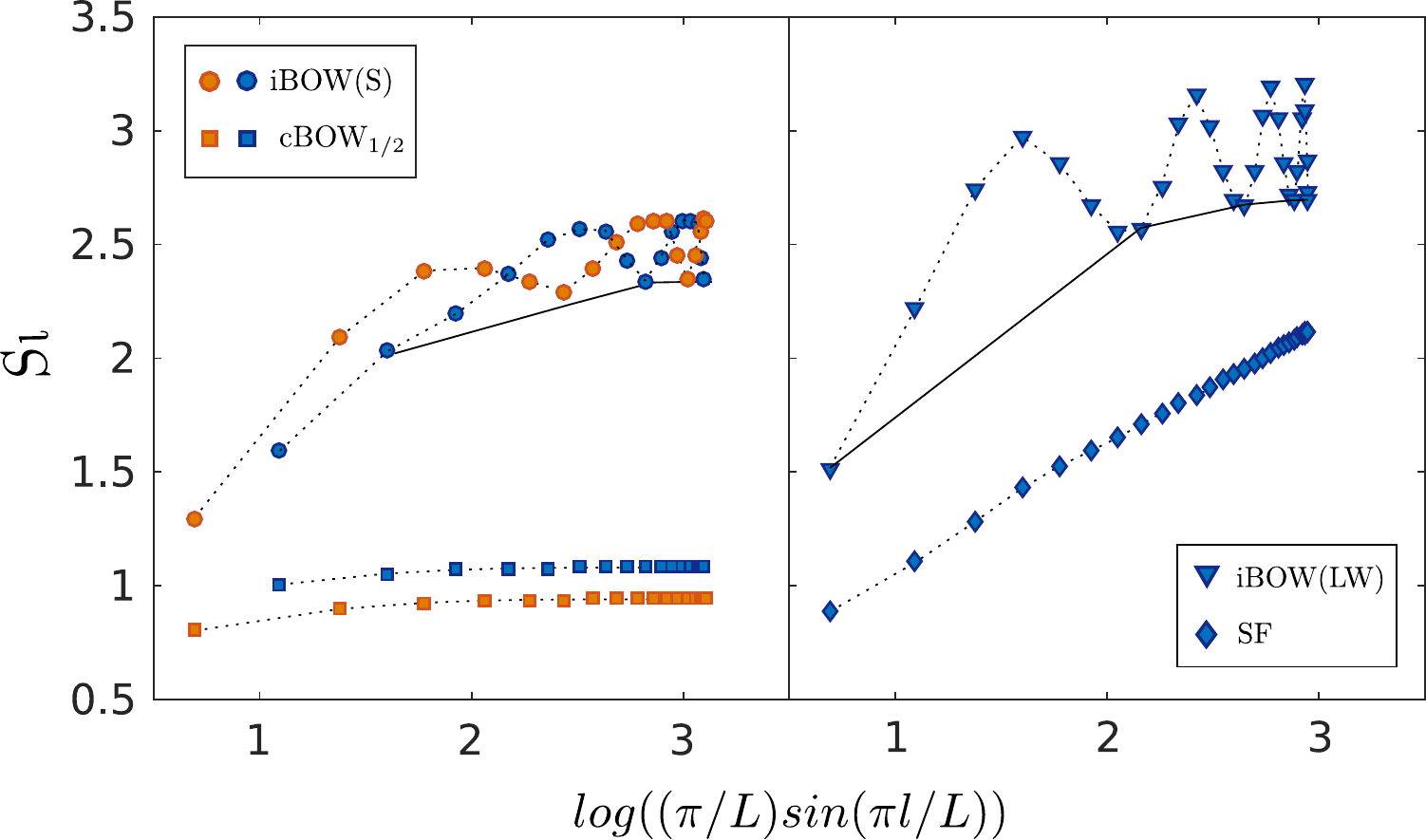}
\caption{\label{fig:entanglement}Entanglement entropy $S_l$, including finite size corrections \cite{calabrese}, in terms of the block size $l$. We show the scaling behavior for different phases, the latter being linear only in the SF case, signaling that this phase is gapless. The spatial inhomogeneities in the ground state influence the scaling. A meaningful analysis is done by increasing the size of the block $l$ by the system's unit cell size. This is depicted by a solid black line for the iBOW phases. In the left panel, odd and even values are represented in different colors.}
\end{figure}

All the BOW phases we observe are gapped. This can be seen by calculating the scaling of the entanglement entropy between a block of size $l$ and the rest of the system, $S_l = -\mathrm{Tr}\,\left(\rho_l\,\mathrm{log}\,\rho_l\right)$, where $\rho_l=Tr_{L-l}|\Psi\rangle\langle\Psi|$ is the reduced density matrix corresponding to the block. For one-dimensional systems, gapped phases follow an area law \cite{area_law}, and $S_l$ saturates when increasing $l$. For gapless phases, however, it grows logarithmically in the block's size \cite{calabrese}. Figure~\ref{fig:entanglement} shows the scaling of the entanglement entropy and indicates that only the SF phase is gapless. The resutls show that the iBOW phases are gapped, although the gap is probably very small. A definitive statement for these phases may require more precise calculations.

\subsection{Superfluid Correlations}

\begin{figure}[t]
  \centering
  \includegraphics[width=1.0\linewidth]{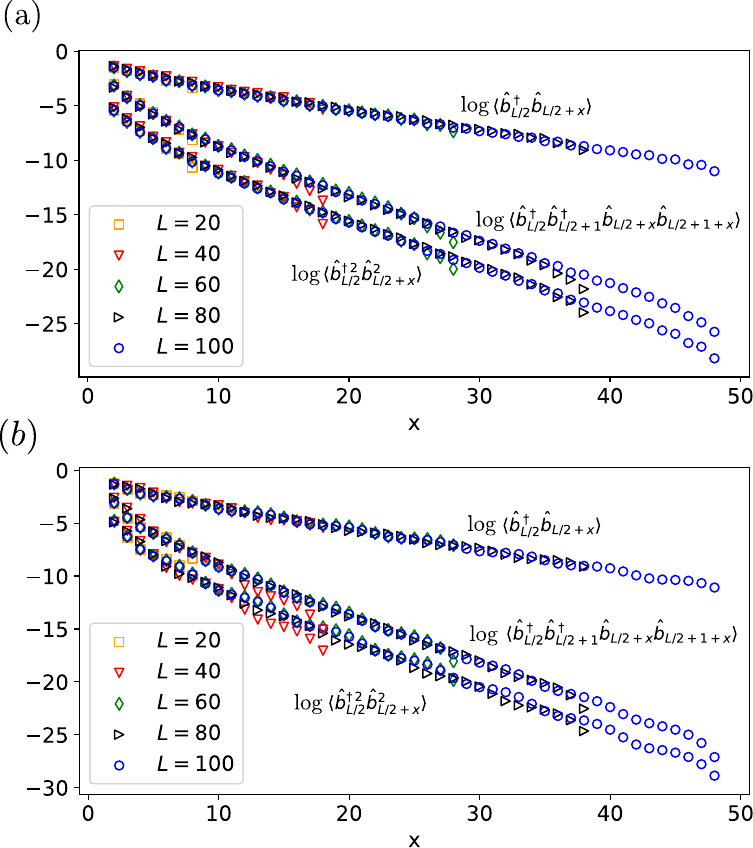}
\caption{\label{fig:correlations}In the figure, we show the logarithm of the superfluid ($\langle\hat{b}^{\dagger}_{L/2}\hat{b}_{L/2+x}\rangle$), on-site pair superfluid ($\langle\hat{b}^{\dagger,2}_{L/2}\hat{b}^2_{L/2+x}\rangle$) and inter-site pair superfluid ($\langle\hat{b}^{\dagger}_{L/2}\hat{b}^{\dagger}_{L/2+1}\hat{b}_{L/2+x}\hat{b}_{L/2+1+x}\rangle$) correlations for a solitonic (a) and a long-wavelength (b) iBOW phase, with $\rho=0.55$ and $\rho=0.85$, respectively. We present the results for different system sizes. The data collapse into straight lines in all the cases, showing that the decay is exponential.}
\end{figure}

None of the BOW phases present quasi-long-range off-diagonal order. In the case of a solitonic and a long-wavelength iBOW phase, we calculated the decay of superfluid ($\langle\hat{b}^{\dagger}_{L/2}\hat{b}_{L/2+x}\rangle$), on-site pair superfluid ($\langle\hat{b}^{\dagger,2}_{L/2}\hat{b}^2_{L/2+x}\rangle$) and inter-site pair superfluid ($\langle\hat{b}^{\dagger}_{L/2}\hat{b}^{\dagger}_{L/2+1}\hat{b}_{L/2+x}\hat{b}_{L/2+1+x}\rangle$) correlations. Figure~\ref{fig:correlations} shows that, in all the cases, the decay is exponential.

\subsection{Soliton - Superfluid Phase Transition}

\begin{figure}[t]
  \centering
  \includegraphics[width=1.0\linewidth]{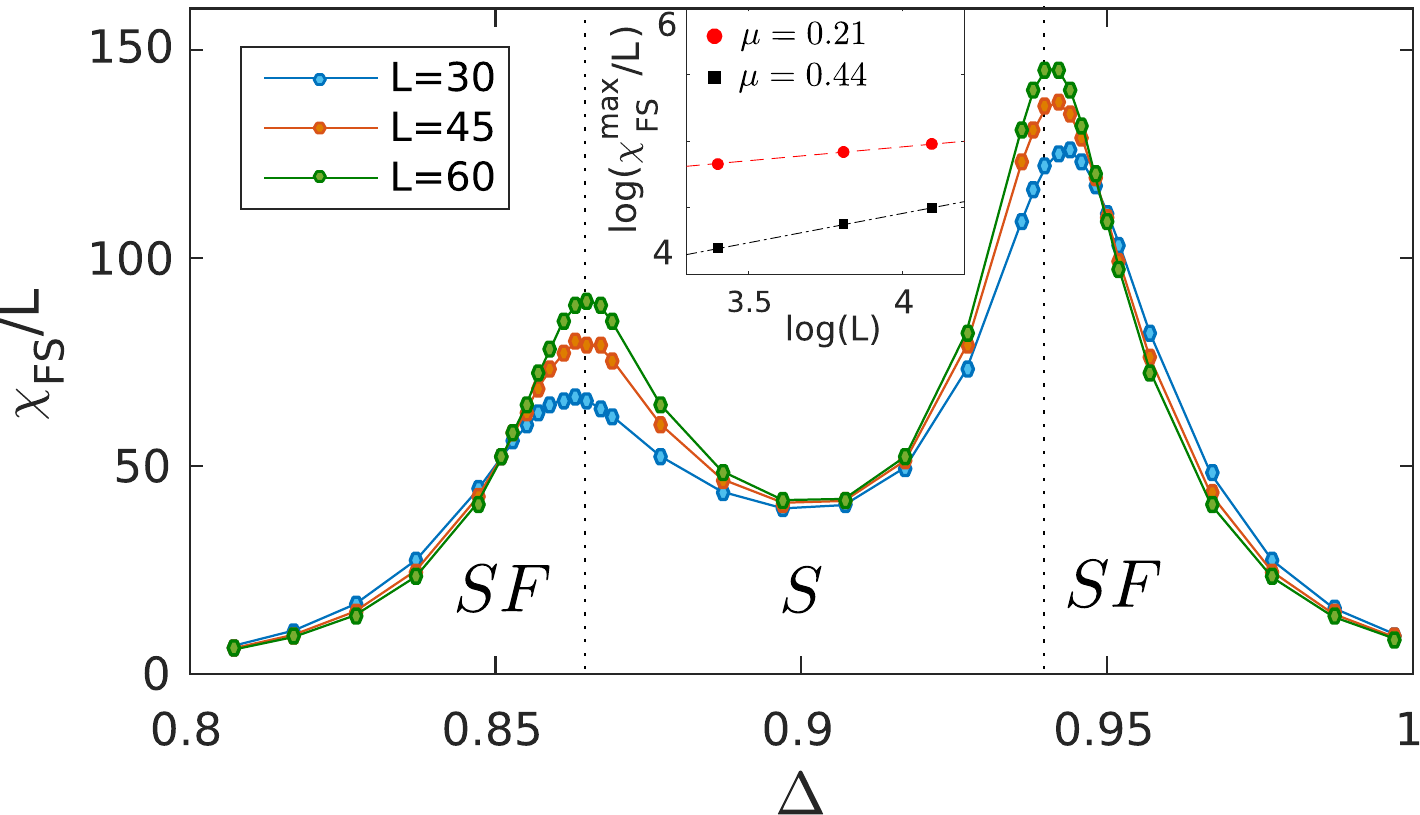}
\caption{\label{fig:suscep}The exact location of the critical points (dotted lines) between the SF and S phases is found using the fidelity susceptibility $\chi_{FS}$. In the figure, $\chi_{FS}$ is represented in terms of $\Delta$ for different system sizes. The critical points are located by extrapolating the positions of the peaks, where $\chi_{FS}$ grows algebraically with the system size, $\chi_{FS}/L\sim L^\mu$. In the inset, the scaling of the value of $\chi_{FS}$ at the peak is represented for the left (black) and right (red) ones. The critical coefficients $\mu$ are obtained by fitting it to a line.}
\end{figure}

The commensurate Bond Order Wave phases can be easily characterized by means of the compressibility, allowing us to locate the transitions between these phases and the soliton and superfluid ones for finite system sizes. Distinguishing between these two, however, is more complicated, since both of them are compressible.  The solitonic and superfluid phases can still be qualitatively distinguished using the structure factor as an order parameter, since the translational invariance symmetry is spontaneously broken in the solitonic phase. However, the exact location of the corresponding critical points is challenging for finite sizes. For this reason, we use the fidelity susceptibility $\chi_{FS}$ \cite{fidelity_susceptibility} to find the critical points in the thermodynamic limit. This quantity can be calculated using the following expression,
\begin{equation}
\chi_{\text{FS}}(\Delta)=\lim_{\delta\to 0}\frac{-2\,\mathrm{log}\,|\langle\Psi_0(\Delta)|\Psi_0(\Delta+\delta)\rangle|}{\delta^2}\,.
\end{equation}
The fideliy susceptibility $\chi_{\text{FS}}$ shows a clear peak near a quantum phase transition, even for small systems. Figure~\ref{fig:suscep} shows $\chi_{\text{FS}}$ in terms of $\Delta$ for different system sizes and a fixed density of $\rho=0.733$, corresponding to a vertical cut in the phase diagram of the system (Fig. 5 in the main text). The critical points are found at $\Delta_1=0.865$ and $\Delta_2=0.940$, by extrapolating the position of the peaks in the thermodynamic limit. In those points, the value of $\chi_{FS}$ grows algebraically as $\chi_{FS}/L\sim L^\mu$ (inset). The critical exponents are different in the two transitions.

\section{Experimental Realization}

\begin{figure}[t]
  \centering
  \includegraphics[width=1.0\linewidth]{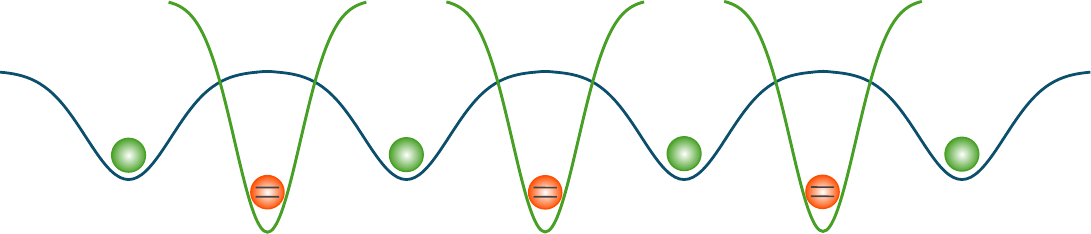}
\caption{\label{fig:experiment}In the figure, we represent the experimental setup used to implement the spin-boson Hamiltonian (Eq. 1 in the main text). The two types of particles (bosons and spins) are trapped in two different optical lattices, located such that the minima of one them lies between two minima of the other one. The first lattice trap bosonic atoms (green in the figure). In the second one, one neutral or charged atom with two internal degrees of freedom is confined (orange). The latter is sufficiently deep to keep the atoms from moving.}
\end{figure}

To implement the boson-spin Hamiltonian (1) of the main text, we consider first a gas of ultracold bosonic atoms trapped in an optical lattice, described by the standard Bose-Hubbard Hamiltonian. A second optical lattice, trapping either neutral or charge atoms, is introduced, placing its minima on the links of the first lattice (Fig.~\ref{fig:experiment}). The atoms trapped in the second lattice have two internal degrees of freedom, and the potential is deep enough to confine them on each minima. The latter represent the spin degrees of freedom of our model when just one atom is loaded on each well. As shown in \cite{atom_ion_zoller,ions_atoms_2,atom_ion_many_body,state_dependent_hopping}, in this situation, the hopping of the moving particles between two neighbor sites is influenced by the internal state of the corresponding spin, giving rise to the following Hamiltonian,
\begin{equation}
\label{eq:bosonic_ham}
\hat{H}_\mathrm{B}=-\sum_{i} \hat{t}_{i}\left(\hat{b}^\dagger_i\hat{b}^{\vphantom{\dagger}}_{i+1}+\mathrm{H.c.}\right)+\frac{1}{2}\sum_i\hat{U}_i\hat{n}_i(\hat{n}_i-1)\,,
\end{equation}
with
\begin{equation}
\hat{t}_{i} = t_\uparrow\ket{\uparrow_{i}}\bra{\uparrow_{i}} + t_\downarrow\ket{\downarrow_{i}}\bra{\downarrow_{i}},
\end{equation}
where $\ket{\uparrow_{i}}$ and $\ket{\downarrow_{i}}$ are the up and down states of a spin on the link $(i,i+1)$, respectively. By rewriting the parameters as $t_\uparrow=t+\alpha$ and $t_\downarrow=t-\alpha$, we obtain the desired boson-spin interaction term,
\begin{equation}
\hat{H}_{\text{BS}}=-\alpha\left(\hat{b}^\dagger_i\hat{\sigma}^z_{i}\hat{b}^{\vphantom{\dagger}}_{i+1}+\text{h.c.}\right)
\end{equation}

The boson-boson interaction is also influence by the presence of the impurity, however, the dependence of $\hat{U}_i$ with the internal state of the impurity is weak \cite{ions_atoms_2}. Therefore, we consider $\hat{U}_i\approx U$.

Consider now the spin dynamics,
\begin{equation}
\hat{H}_\mathrm{S}=\frac{\Delta}{2}\sum_{i}\hat{\sigma}^z_{i}+\beta\sum_{i}\hat{\sigma}^x_{i}\,,
\end{equation}
which is composed solely of on-site terms. The first term is the energy difference between the two spin states (internal states of the impurity) and can be tuned using an external magnetic field. The second term corresponds to a spin flipping, and can be is enforce using laser-assisted transitions between the two states. This approach works both when the impurity correspond to neutral atoms or ions.

\section*{Numerical Method}

The study of a model composed of softcore bosons and spins is computationally challenging due to the high local dimension of the physical system. Here, we use the density-matrix renormalization group algorithm \cite{dmrg}. In order to reduce the numerical effort we implement to types of approximations: we truncate the maximum number of bosons per site to $n_0=2$ and the bond dimension to $D=40$. In the following, we justify we these approximations are meaningful.

\subsection*{Local Number of Bosons}

\begin{figure}[t]
  \centering
  \includegraphics[width=0.9\linewidth]{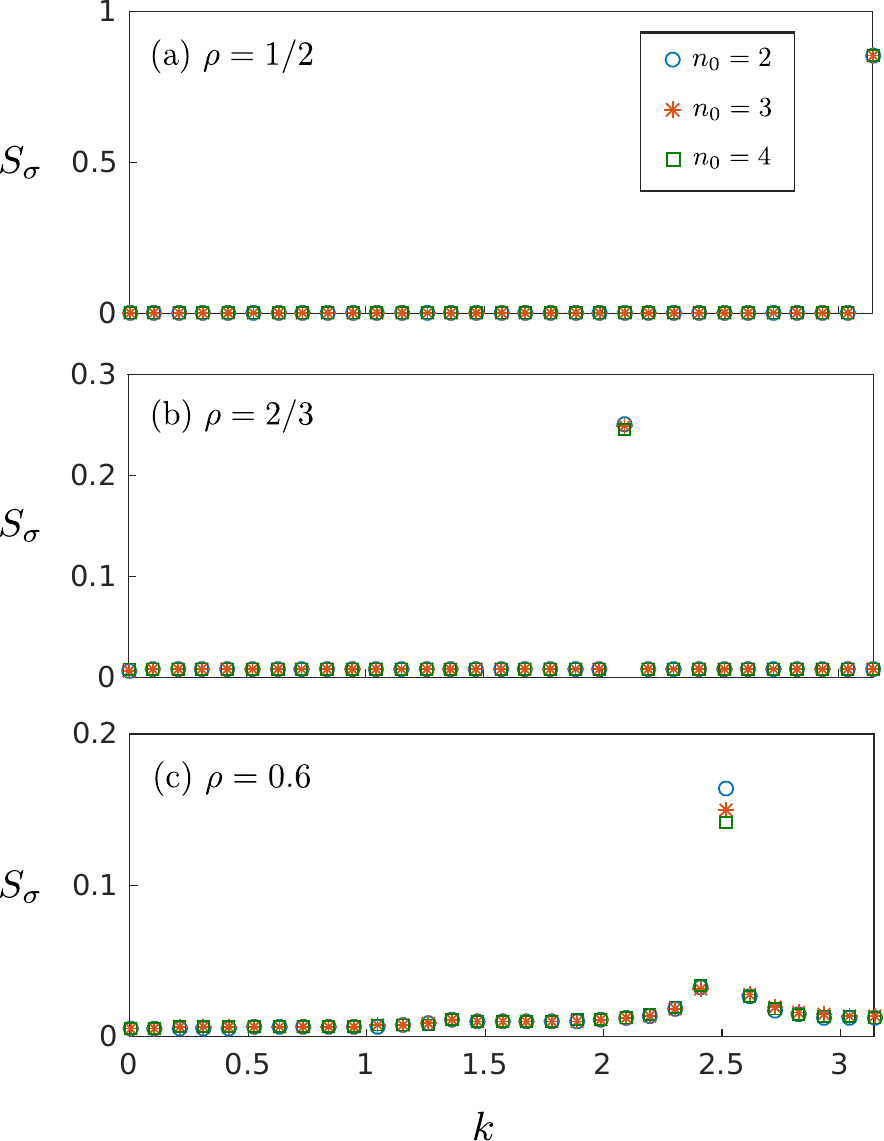}
\caption{\label{fig:n0}Structure factor $S_\sigma$ in terms of the momentum $k$ for states in different BOW phases, for different values of the maximum number of bosons per site $n_0$.}
\end{figure}

High bosonic occupation is suppressed in the system since we are considering low densities, $0<\rho\le 1$, and strong interactions, $U/t=10$. We have checked that all the phases we describe are stable when the value of $n_0$ is increased. As an example, we show in Figure~\ref{fig:n0} the structure factor for states in different BOW phases. We can observe how, although the height of the peak might vary a bit in the solitonic phase, the peak itself persist, and so do the rest of the properties of the phase. In particular, the expectation value of the bosonic occupation in real space do not change appreciably. This situation is analogous for the rest of the phases presented in this work.

\subsection*{Bond Dimension}

\begin{figure}[t]
  \centering
  \includegraphics[width=1.0\linewidth]{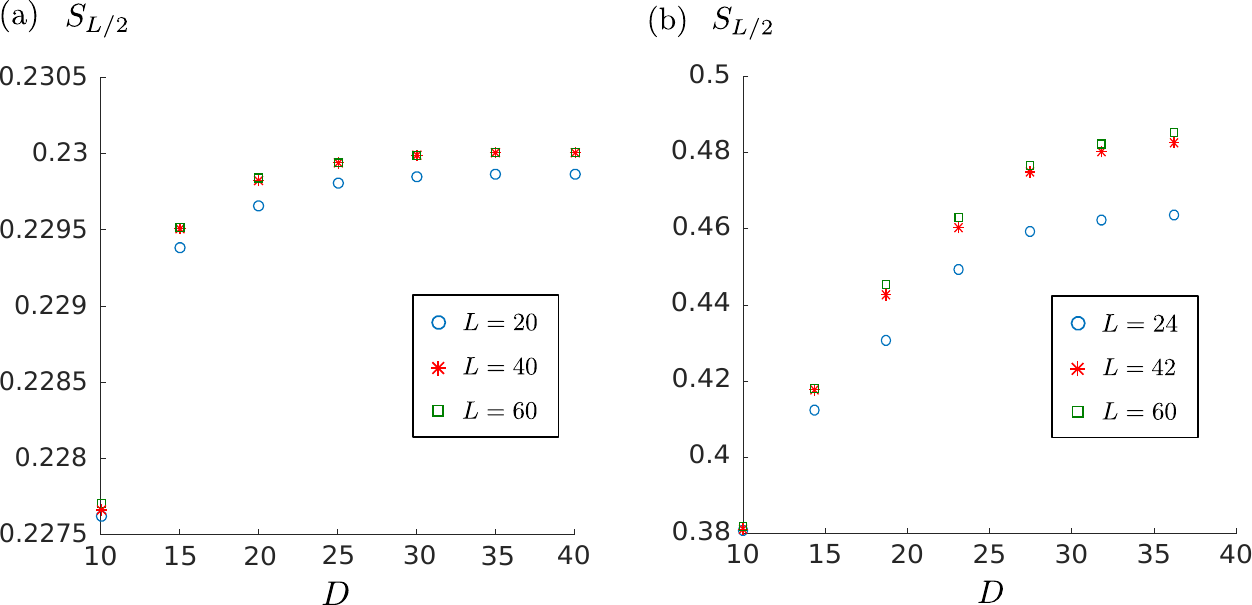}
\caption{\label{fig:entropy}Entanglement entropy calculated at the middle of the chain, $S_{L/2}$ in terms of the bond dimension $D$ for different system sizes $L$. (a) and (b) correspond to the BOW$_{1/2}$ and BOW$_{2/3}$, respectively.}
\end{figure}

The area law for the entanglement entropy \cite{area_law} guarantees that we can approximate with arbitrary precision in one-dimensional gapped states using an MPS with finite bond dimension D \cite{dmrg}. Figure~\ref{fig:entropy} shows the entanglement entropy for half of the chain in terms of $D$ for different system sizes, both for a state in the BOW$_{1/2}$ and BOW$_{2/3}$ phases. In the first case (a), the entropy saturates both in $D$ and $L$. The former means that a bond dimension of $D=40$ is enough to totally describe the state and all its possible correlations. The latter implies that the bulk properties of the system are not affected by finite-size effects (due to the finite correlation length) for a system size of $L=60$, and it is equivalent to the bulk in the thermodynamic limit. In the second case (b), the gap of the system is smaller. For this reason, a complete saturation in terms of $D$ is not obtained. However, one can see how the entropy does not change much when increasing the bond dimension, meaning that it is almost saturated.

\bibliographystyle{apsrev4-1}
\bibliography{supp_bibliography_v2}